\documentclass[twocolumn,showpacs,preprintnumbers,amsmath,amssymb]{revtex4}

\usepackage{graphicx}
\usepackage{dcolumn}
\usepackage{bm}

\def\raamno{$R_{1-x}A_x$MnO$_3$}
\def\raamnmo{$R_{1-y}A_y$Mn$_{1-x}M_x$O$_3$}
\def\lsmbo{La$_{1-y}$Sr$_y$Mn$_{1-x}$$M_x$O$_3$}
\def\lsmzno{La$_{1-y}$Sr$_y$Mn$_{1-x}$Zn$_x$O$_3$}
\def\lsmcuo{La$_{1-y}$Sr$_y$Mn$_{1-x}$Cu$_x$O$_3$}

\def\lsmcro{La$_{1-y}$Sr$_y$Mn$_{1-x}$Cr$_x$O$_3$}
\def\lsmcoo{La$_{1-y}$Sr$_y$Mn$_{1-x}$Co$_x$O$_3$}
\def\lsmo{La$_{0.7}$Sr$_{0.3}$MnO$_3$}
\def\lamno{La$_{0.949}$Mn$_{0.949}$O$_3$}
\def\cm{\,{\rm cm^{-1}}}
\def\PP{$e_{\rm i}\parallel e_{\rm s}$}
\def\CP{$e_{\rm i}\perp e_{\rm s}$}
\def\tc{$T_{\rm c}$}
\def\etal{{\it et~al.}}
\def\K{\,{\rm K}}
\def\cm{\,{\rm cm^{-1}}}
\def\mo{{\it M}--O}

\def\bo{{\it B}--O}

\def\nic #1{}

\begin{document}

\preprint{}

\title{%
\boldmath
Raman and infrared studies  of 
\lsmbo\ ($M$=Cr, Co, Cu, Zn, Sc or Ga): Oxygen disorder and local vibrational modes%
\unboldmath
}

\def\spainaff{\affiliation{%
Materials Science Institute, University of Valencia, P.O. Box 22085, 46071 Valencia, Spain 
}%
}

\def\czaff{\affiliation{%
Institute of Condensed Matter Physics, Faculty of Science, Masaryk
University, Kotl\'a\v{r}sk\'a 2, CZ--61137 Brno, Czech republic
}%
}

\author{A. Dubroka} 
\email{dubroka@physics.muni.cz}
\spainaff
\czaff
\author{J. Huml\'{\i}\v{c}ek}
\czaff
\author{M.V.~Abrashev}
\affiliation{Faculty of Physics, University of Sofia, BG--1164 Sofia, Bulgaria }

\author{Z.V. Popovi\'{c}}%
\altaffiliation[Permanent address: ]{Center for Solid State Physics and New Materials, Institute of Physics, P.O.
Box 68, 11080 Belgrade/Zemun, Serbia}
\author{F. Sapi\~{n}a}
\spainaff
\author{A. Cantarero}
\spainaff

\date{\today}

\begin{abstract} We present results of our study of polarized Raman scattering
	and infrared reflectivity of rhombohedral ceramic \lsmbo\ manganites in
	the temperature range between 77 and 320\,K.  In our samples, a part of
	the Mn atoms is substituted by $M$ = Cr, Co, Cu, Zn, Sc, or Ga with $x$
	in the range 0--0.1.  The hole concentration was kept at the optimal
	value of about $32\%$ by tuning the Sr content $y$.  We have monitored
	distortions of the oxygen sublattice by the presence of broad bands in
	the Raman spectra, the increase of dc resistivity extracted from the
	infrared reflectivity, and the change of the critical temperature of
	the ferromagnetic transition.  Our results support the idea that  these
	properties are mainly determined by the radius of the substituent ion,
	its electronic and magnetic structure playing only a minor role.
	Furthermore, the Raman spectra exhibit an additional $A_g$-like high
	frequency mode attributed to the local breathing vibration of oxygens
	surrounding the substituent ion.  Its frequency and intensity strongly
	depend on the type of the substituent.  In the Co-substituted sample,
	the mode anomalously softens when going from 300 to 77\,K.  The
	frequency of the bulk $A_{1g}$ mode depends	linearly on the angle
	of the rhombohedral distortion. 
	
\end{abstract}

\maketitle

\section{Introduction} Recently, considerable interest has been paid to
\raamno\ manganites that are characterized by a strong interplay of spin,
charge and lattice degrees of freedom~\cite{Salamon}.  The research is
motivated both by fundamental physical questions and by potential applications
of the colossal magnetoresistance (CMR) appearing near the temperature of the
ferromagnetic transition \tc.  Much effort has also been devoted to the
compounds \raamnmo\ substituted at the Mn site.  Substitutions of the Mn ion,
which is at the heart of the double exchange (DE) interaction, may contribute
to the knowledge of the basic mechanism of CMR and also allows one to tune the
characteristics of the compounds such as \tc.  Many researchers have addressed
the transport and magnetic properties~\cite{Ghosh,Ahn, Fadli}, covering a wide
range of materials with $R$ = La, $A$ = Ca, Sr, Ba and $M$ = Al, Co, Cr, Cu,
Fe, Ni, Sc, Ti, and Zn.  The substitutions lead to a reduction of \tc\ and
magnetization, an increase of resistivity and an enhancement of the
magnetoresistance.  In explaining these effects, a significant role has been
attributed to the local elastic stress produced by the substituent and to its
electronic and magnetic character. Only a few studies concerned the optical
properties~\cite{Filho,MarziRaman,MarziInfra}. 

In this paper we present results of polarized Raman scattering and an infrared
reflectance study of ferromagnetic rhombohedral \lsmbo\ compounds ({\it M} =
Cr, Co, Cu, Zn, Sc or Ga) for several concentrations of the substituent ($x$ =
0.02, 0.04, 0.06, 0.08, 0.1 for Cr, Co, Cu and Zn, $x$ = 0.1 for Sc and $x$ =
0.08 for Ga).  In the compound, the substituents influence locally their
surroundings and the average concentration of holes in the sample.  In order to
see more clearly the former effect, the concentration of holes has been kept
approximately at 32\% by adjusting the Sr content (see Sec.~\ref{experiment}).
The presence or absence of broad bands around 500 and $620\cm$ in Raman spectra
allows us to distinguish between strong and weak distortions of the oxygen
sublattice.  The distortions manifest themselves also in an increase of
resistivity extracted from the infrared reflectivity.  We discuss these effects
and the variations of \tc\ in terms of the radii of the substituents inferred
from the x-ray data.  We discuss in detail the properties of an additional
$A_{g}$-like high-frequency peak found in Raman spectra.  We investigate the
relation of the frequency of the bulk $A_{1g}$ mode to the angle of the
rhombohedral distortion.

The paper is organized as follows. In Sec.~\ref{experiment} we discuss the
details of the experiments. In Sec.~\ref{XRayTc} we comment on the results of
x-ray analysis and variations of \tc.  The main body of the paper
(Sec.~\ref{Raman}) is devoted to the Raman data. The results of the infrared
studies are presented in Sec.~\ref{infra} and a summary in
Sec.~\ref{conclusion}.

\section{Samples and Experiment} \label{experiment} Polycrystalline, single
phase \lsmbo\ powders have been prepared by thermal decomposition of precursors
obtained by a freeze-drying of acetic acid solutions.  This soft procedure makes
possible a strict control of stoichiometry, and allows us to keep the
concentration of vacancies at the La and Mn site practically negligible.
Pellets were prepared by pressing and sintering the powder at $1150\ ^\circ$C
for 48~h. The values of the parameters $x$ and $y$ were selected according
to the valence of the substituent in order to maintain the constant proportion
of Mn$^{4+}$/Mn$^{3+}$ ions of about 32\%, see Table~\ref{TableXY}. The
valencies of the substituents were chosen as follows: Sc$^{3+}$ (the electron
configuration 3$d^0$), Ga$^{3+}$ (3$d^{10}$) and Zn$^{2+}$ (3$d^{10}$) ions are
known to be the only possible states; the valencies of Cr$^{3+}$ (3$d^{3}$),
Co$^{3+}$ (3$d^{6}$) ions are likely due to the charge neutrality condition.
We have further assumed the valence state 2+ of Cu (3$d^{9}$) because the 3+
state is rare.  However, it is not excluded that the actual valency of Cu is
not integer and exceeds 2+.  X-ray powder diffraction patterns were completely
indexed with rhombohedral perovskite cells. For further details of the sample
preparation and results of the diffraction analysis see Ref.~\cite{Fadli}.
Table~\ref{TableTc} presents the values of \tc\ of our samples.  The \lamno\
compound, which was prepared by the same procedure, was found to be a
noncollinear ferromagnet with $T_{\rm c}=145\K$~\cite{Vergara}. 

\begin{table}[t]
\begin{tabular}{p{2cm}p{2cm}p{2cm}p{2cm}}
\hline
\hline
\multicolumn{4}{c}{\vrule height 10pt width 0cm  \lsmbo\ }\\ 
\multicolumn{2}{l}{$M=$ Cu or Zn } & \multicolumn{2}{l}{$M=$ Cr, Co, Ga or Sc}\\
\hline
$x$  &   $y$ &  $x$ & $y$   \\
\hline
\vrule height 10pt width 0cm 
0    & 0.30   \\
0.02 & 0.274 & 0.02 & 0.294 \\
0.04 & 0.248 & 0.04 & 0.288 \\
0.06 & 0.222 & 0.06 & 0.282 \\
0.08 & 0.196 & 0.08 & 0.276 \\
0.10 & 0.170 & 0.10 & 0.270 \\
\hline
\hline
\end{tabular}
\caption{Compositions of the \lsmbo\ samples.}
\label{TableXY}
\end{table}

The Raman spectra were measured with a triple Jobin-Yvon 64000 spectrometer
equipped with a liquid nitrogen cooled charge coupled device detector; the
resolution was set to $2.7\cm$. The 488\,nm Ar--Kr laser line was focused on a
spot of a typical size of about 0.1\,mm by a lens with the focal length of
10~cm, and collected in the pseudobackward scattering geometry. The power of
the laser at the sample was kept below 18\,mW. From the frequency shift of the
soft mode around $200\cm$ we have estimated that the overheating of the spot on
the sample was less than 10\,K.  Samples were cooled using a continuous flow
cryostat.

\begin{table}[b]
\def\collen{0.8cm}
\begin{tabular}{ p{\collen} p{\collen} p{\collen} p{\collen} p{\collen} p{\collen} p{\collen} }
\hline
	\hline
& \multicolumn{6}{c}{$x$} 
\\
\cline{2-7}
\vrule height 10pt width 0cm 
$M$  & 0 & 0.02 & 0.04 & 0.06 & 0.08 & 0.10 \\
\hline
\vrule height 10pt width 0cm 
Cr & 372 & 363 & 351 & 340 & 327 & 316  \\
Co &     & 355 & 345 & 331 & 312 & 300  \\
Cu &     & 358 & 331 & 308 & 274 & 236  \\
Zn &     & 350 & 326 & 284 & 233 & 179  \\
Sc &     &     &     &     &     & 195  \\
Ga &     &     &     &     & 320 & 310  \\
\hline
\hline
\end{tabular}
\caption{Values of \tc\ (K) of our \lsmbo\ samples,
taken from Ref.~\cite{Fadli}.}
\label{TableTc}
\end{table}

Raman signals from polycrystalline samples have been collected in the macro
regime from a number of randomly oriented microcrystals because the spot of the
light is much larger than the typical size of the microcrystals of about
$1\,\mu$m.  We have performed polarized measurements in two scattering geometries
with parallel ($e_{\rm i}\parallel e_{\rm s}$) and crossed ($e_{\rm i}\perp
e_{\rm s}$) polarizations  of the incident and the scattered light,
respectively.  Although we are not able to extract individual components of the
Raman tensor, these two geometries have different selection rules for the Raman
lines, providing more information than unpolarized measurements. To get rid of
the trivial temperature dependence,  all spectra have been divided  by the
Bose--Einstein occupation number
$n(\omega)+1=1/[1-\exp(-\frac{\hbar\omega}{k_{\rm B}T})]$.

The infrared reflectivity was measured in the range 80--7000$\cm$ using a
Bruker IFS 66v/S spectrometer.  The samples were cooled using a Janis
ST--100--FTIR continuous flow cryostat.

\section{Results and Discussion} 
\label{results} 

\boldmath
\subsection{X-ray data and \tc} 
\unboldmath
\label{XRayTc}

\begin{figure}[t]
\vspace{-10mm}
	\includegraphics[width=8.6cm]{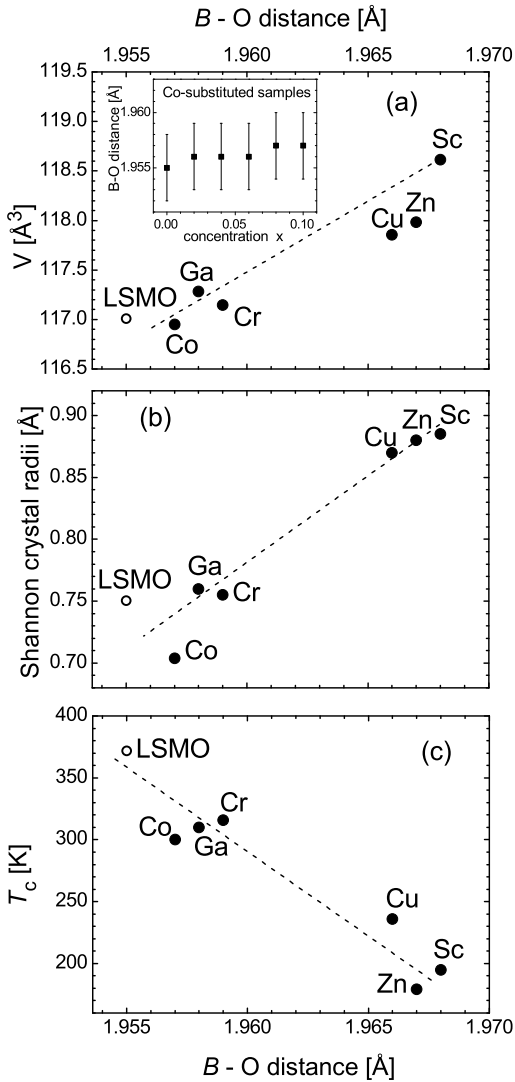}
\vspace{-5mm}
	\caption{Volume of the elementary cell (a),  Shannon radius
	of the substituents (b), and  \tc\ (c), versus the \bo\ distance of the \lsmbo, $x=0.10$
	samples and the parent $x=0$ (LSMO) sample. 
	X-ray data are taken from Ref.~\cite{Fadli}. The dashed lines are shown to guide the
	eye. The inset of the part (a) displays the \bo\ distance of the Co-substituted samples.}
\label{XRay}
\end{figure}

An insight into the differences among the substituents can be obtained by
analyzing the x-ray data~\cite{Fadli}.  A particularly important information is
provided by the \bo\ distance, i.e., the mean distance between the central ion
in the perovskite {\it AB}O$_3$ structure and neighbouring oxygens.  In the case of
the 10\% substituted samples, this quantity represents the average of nine
Mn--O distances and one {\it M}--O distance.  Since the average valency of the
Mn ions is kept constant and their radius is thus fixed, the \bo\ distance
should depend on the radii of the {\it M} ions. Figure~\ref{XRay}(a) displays
the volume of the elementary cell versus the \bo\ distance. It is apparent that
the compounds can be sorted into two groups according to the \bo\ distance: (i)
close to that of the parent compound (Cr, Co and Ga), and (ii) significantly
larger (Sc, Zn and Cu). This indicates that the radii of the substituents
within the former (latter) group are similar (larger) to the radius of the Mn
ion.  This differentiation can also be noted in the variations of the volume
of the elementary cell. The volume, however, is a quite complex quantity. For
example, a decrease of the volume of the La$_{1-y}$Sr$_y$MnO$_3$ system with
increasing concentration of Sr (Ref.~\cite{Chmaissem}) contradicts a simple
expectation based only on the Shannon radii of La and Sr (that of Sr is
larger). The observed trend is probably caused by the increase of the
concentration of the Mn$^{4+}$ ions that are smaller than the Mn$^{3+}$ ions.
In our samples, where the concentration of holes is kept constant, the
variations of the volume can thus be expected to follow predominantly those of
the substituent radii.  However, the volume of Cu- and Zn-substituted samples
is lower than that of the Sc-substituted sample due to a higher concentration
of Sr in the latter.  The inset of Fig.~\ref{XRay}(a) displays the \bo\
distance of Co-substituted samples as the function of the concentration $x$
including error bars corresponding to the standard deviation of 0.003\,\AA\
resulting from X-ray analysis~\cite{Fadli}. Obviously, the dispersion of
experimental points is much lower than 0.003\,\AA, our estimate is about
0.001\,\AA.  Because of this difference, we do not display the error bars of
the \bo\ distance in our figures.  Since the present paper is focused mainly on the
distinction between the groups of samples (i) and (ii), the conclusions will
not be influenced even by the error margin of 0.003\,\AA.  A similar discrepancy
between the nominal standard deviation and the real dispersion of data points
is apparent in Fig.~1 of Ref.~\cite{Blasco}.

The values of the \bo\ distance are compared with the Shannon
radii~\cite{Shannon} in Fig.~\ref{XRay}(b); the latter correspond to the
valence of 3+ for Sc, Cr, Co and Ga ions, and 2+ for Cu and Zn ions. For Co, we
have used the value of the intermediate spin state that is estimated to be
about 3\% larger than that of the low spin state~\cite{Radaelli}.  The radius
of Mn was calculated as the weighted average of those of the Mn$^{3+}$ and
Mn$^{4+}$ ions, for the mean oxidation state of 3.3.  The Shannon radius
correlates reasonably well with the \bo\ distance, supporting the relevance of
the representation of the substituent radii by the \bo\ distance. Moreover, the
\bo\ distance seems to be essentially independent of the Sr concentration. For
example, the values for the Zn and Sc points corresponding to the Sr content of
0.17 and 0.27, respectively, are almost the same.

Figure~\ref{XRay}(c) displays $T_{\rm c}$ versus the \bo\ distance. A
pronounced reduction of \tc\ occurs for  Cu-, Zn- and Sc-substituted samples,
where the radius of the substituent is larger than that of the Mn ion. This
correlation has been attributed to a modification of the Mn--O distances around
the substitutional ion~\cite{Ghosh}, leading to changes of the energies of the
$e_g$ orbitals [similar to those of the Jahn--Teller (JT) effect], and to a
localization of the itinerant electrons that reduces the DE interaction and
\tc.

\subsection{Raman scattering} 
\label{Raman}

\label{assignment} Our samples crystallize in rhombohedral structure, space
group $D^6_{3d}, Z=2$.  This structure can be obtained from the simple cubic
perovskite  by the rotation of the adjacent MnO$_6$ octahedra in the opposite
directions around the [111]$_c$ cubic direction. Of the total of the 20
$\Gamma$-point modes ($A_{1g}+3A_{2g}+2A_{1u}+4A_{2u}+4E_{g}+6E_{u}$), five
($A_{1g}+4E_{g}$) are Raman active, eight ($3A_{2u}+5E_{u}$) are infrared
active, and five ($2A_{1u}+3A_{2g}$) are silent.

\begin{figure}[!t]
\vspace{-10mm}
	\includegraphics[width=8cm]{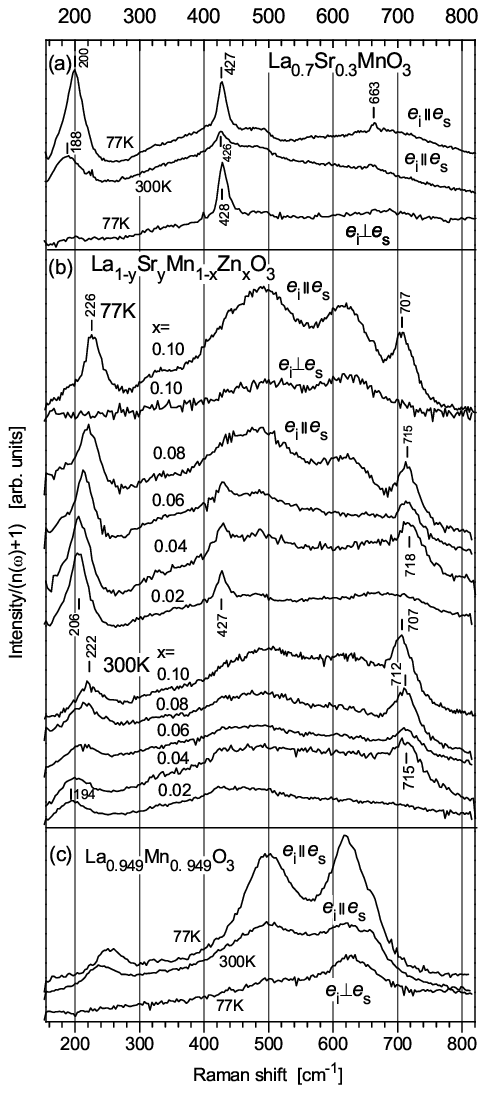}
	\vspace{-5mm}
	\caption{Raman spectra of (a) \lsmo, (b) \lsmzno, and (c) \lamno\ polycrystalline
samples measured for the parallel
polarizations of the incident and the scattered light. The spectra measured for the crossed
polarizations are marked by \CP.}
\label{Zn}
\end{figure}

\begin{figure}[!t] 
\vspace{-10mm}
	\includegraphics[width=8cm]{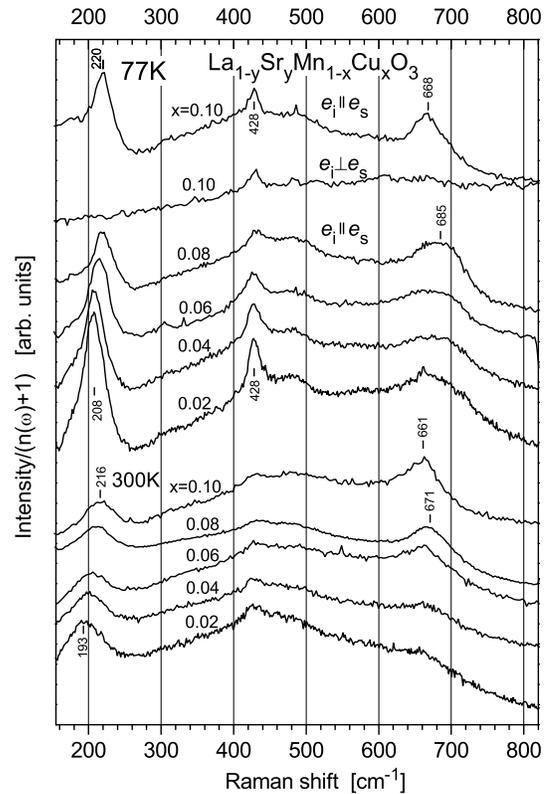}
	\vspace{-5mm}
	\caption{Raman spectra of polycrystalline \lsmcuo\ samples. All spectra were measured
	for parallel polarizations  of the incident and the 
	scattered light except for the spectrum of
	the $x=0.10$ sample, marked by \CP, which was measured for the crossed
	polarizations.} 
\label{Cu} 
\end{figure}

Figure~\ref{Zn}(a) shows our Raman spectra of the parent compound \lsmo\ at
$77$ and 300\K.  The low temperature spectrum measured in the \PP\
polarizations exhibits three lines at $200$, $427$, and $663\cm$. The shell
model calculation of the isostructural rhombohedral LaMnO$_3$ from
Ref.~\cite{Abrashev} predicts the frequency of the $A_{1g}$ mode of $249\cm$
and the frequencies of $E_{g}$ modes at 42, 163, 468, and $646\cm$.  Similarly
to previous studies~\cite{Podobedov,Abrashev,Granado}, we assign the $200\cm$
line  to the $A_{1g}$ mode, and the $427\cm$ line to the $E_{g}$ mode  with the
predicted frequency of $468\cm$.  This assignment is supported by the different
symmetry of the modes that is apparent in the \CP\ spectrum where only the
$E_g$ mode is present. The frequencies of the two modes are close to the
results obtained on a single crystal (199 and 426$\cm$,
respectively~\cite{Granado}).  In all our spectra, the $E_g$ mode has almost
the same frequency, $427\pm2\cm$ at 77\K, and it softens slightly with
increasing temperature. We do not assign the weak band at $663\cm$ to the
manganite because its intensity varies substantially from measurement to
measurement. Similarly to Podobedov~\etal~\cite{Podobedov}, we assume that it
is due to a small volume of manganese oxides, e.g., Mn$_3$O$_4$ has a strong
Raman line at about this frequency~\cite{Liu}.

Figure~\ref{Zn}(b) shows the Raman spectra of the Zn-substituted series. With
increasing Zn concentration, the spectra exhibit the following main features:
the $A_{1g}$ mode hardens, a fairly strong mode appears at about $710\cm$, the
mode at $427\cm$ gradually disappears while two broad bands appear at $500$ and
$620\cm$. The broad bands are discussed in Sec.~\ref{JTbands}. 

\begin{figure}[!t] 
\vspace{-10mm}
	\includegraphics[width=8cm]{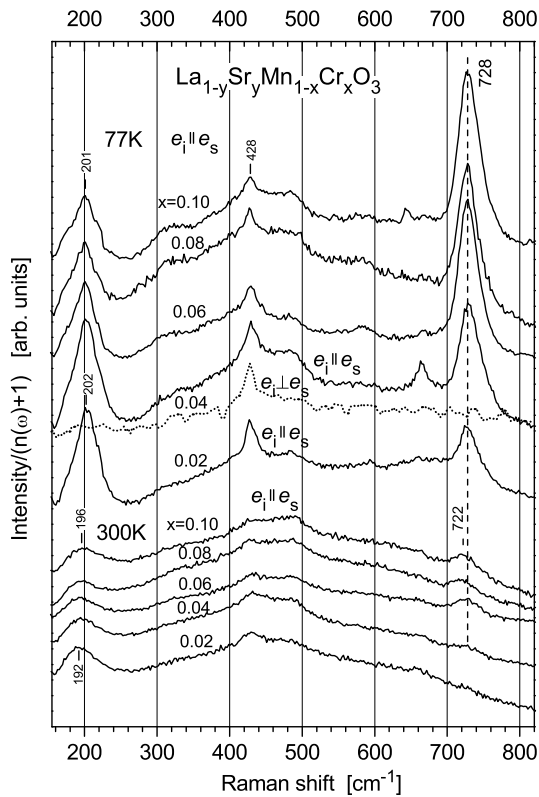}
	\vspace{-5mm}
	\caption{Raman spectra of polycrystalline  \lsmcro\ samples. All spectra were measured
	for parallel polarizations  of the incident and the 
	scattered light except for the spectrum of
	the $x=0.04$ sample, marked by \CP, which was measured for the crossed
	polarizations.} 
\label{Cr} 
\end{figure}

\begin{figure}[!t] 
\vspace{-10mm}
	\includegraphics[width=8cm]{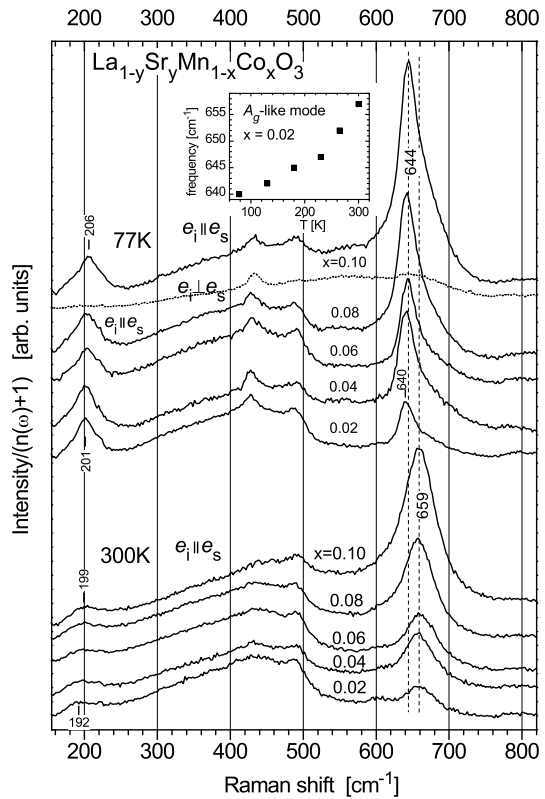}
\vspace{-5mm}
	\caption{Raman spectra of polycrystalline \lsmcoo\ samples. All spectra were measured
for parallel polarizations  of the incident and the scattered light except for one
spectrum of the $x=0.10$ sample, marked by \CP, which was measured for the crossed
polarizations. The inset shows the frequency of the $A_{g}$-like mode of the $x=0.02$ sample
versus temperature.} 
\label{Co} 
\end{figure}

Figures~\ref{Cu}, \ref{Cr}, and \ref{Co} display the Raman spectra of the Cu, Cr,
and Co series, respectively. First, let us focus on the 77\K\ spectra measured
in the \PP\ polarizations. With increasing concentration of the substituent,
the oxygen bending mode at $427\cm$ weakens but remains well resolved even at
the largest concentration of $x=0.10$. However, even at the highest $x$ none of
the spectra exhibits the broad bands such as in the Zn-substituted series.  A
high-frequency mode emerges at about $670\cm$ (Cu), $728\cm$ (Cr) and $644\cm$
(Co), respectively.  The mode is discussed in detail in Sec.~\ref{Aglike}. The
low-frequency $A_{1g}$ mode hardens with increasing concentration of
substituents, which will be discussed in Sec.~\ref{soft}.  We have performed
temperature studies (not shown) between 300 and 77\K\ of several selected
samples (Zn 10\%, Zn 6\%, Cu 10\%, Cu 8\%, Cr 4\%, and Co 2\%). The spectra
exhibit basically a gradual change between 300 and 77\K.

\begin{figure}[!t] 
\vspace{-10mm}
	\includegraphics[width=8cm]{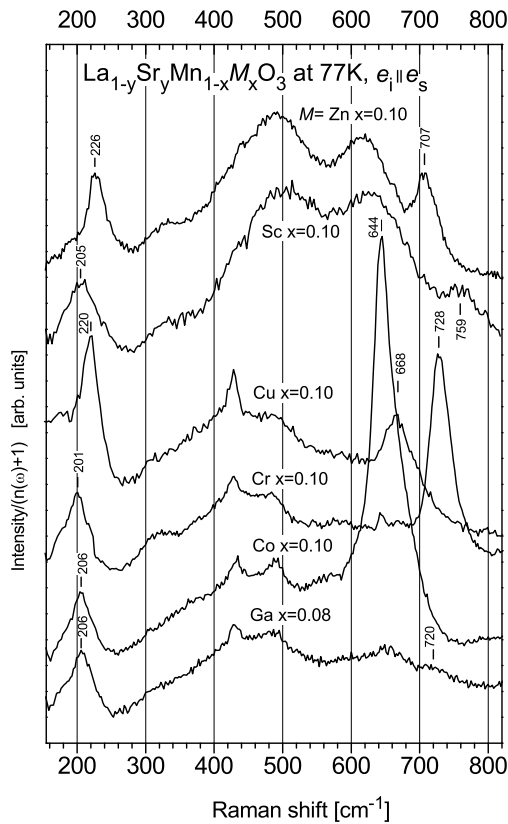}
\vspace{-5mm}
	\caption{Raman spectra of \lsmbo, $x=0.10$, samples (except for $M=$ Ga with
$x=0.08$) measured for parallel polarizations of the incident and the scattered light.} 
\label{prehled10} \end{figure}

Figure~\ref{prehled10} summarizes the 77\K\ spectra of the $x=0.10$ compounds
including the one of Sc, which  exhibits broad bands similar to those of the Zn
sample and a high-frequency peak at $759\cm$. The 77\K\ spectrum of the Ga
$x=0.08$ sample is shown as well.  A Ga-substituted sample with $x=0.10$ was
not available for our optical studies; however, we expect the difference
between the samples with Ga concentration of 0.10 and 0.08 to be rather small
and the absence of this particular composition does not degrade our qualitative
conclusions.  In the spectrum of the Ga sample, there is no pronounced line at
high frequencies; however, it is possible that a high-frequency mode similar to
those of the remaining samples is present as the weak structure around
$720\cm$.

\subsubsection{Jahn-Teller-like bands} \label{JTbands} As far as the broad
bands are concerned, the spectra displayed in Fig.~\ref{prehled10} can be
divided into two groups: that of the Zn- and Sc-substituted samples displaying
them, and that of Cu-, Cr-, Co-, and Ga-substituted samples where they are
absent.  Spectral structures very similar to these broad bands were observed in
the paramagnetic phase of La$_{0.7}$Ca$_{0.3}$MnO$_{3}$~\cite{IlievB} and
La$_{0.8}$Sr$_{0.2}$MnO$_{3}$~\cite{Podobedov2,Bjornson}, disappearing in the
ferromagnetic metallic phase.  These structures (``JT bands'') have been
assigned to the scattering on the oxygen sublattice distorted by the JT
effect~\cite{Iliev}. The JT distortion is not compatible with the rhombohedral
structure where all Mn--O bonds have equal length, and results in a
noncoherently distorted oxygen sublattice. Raman scattering on such structures
is not restricted by selection rules and its intensity is proportional to the
phonon density of states~\cite{Shuker}. The disappearance of the bands in the
metallic phase corresponds to a decrease of JT distortions due to the
delocalization of the $e_{g}$ electrons. 

We suggest that the broad bands in the Zn- and Sc-substituted compounds are
also caused by distortions of the oxygen sublattice.  In these samples,
however, the distortions are primarily induced by the substituents;
consequently, we call the broad bands JT-like bands rather than JT bands.  The
intensity of the JT-like bands obviously increases with the concentration of
the substituents.  The appearance of the JT-like bands correlates with the \bo\
distance (or, equivalently, with the radius of the substitutional ion): they
are present in the spectra of Sc- and Zn-substituted compounds (large radii)
and absent in Cr-, Co-, and Ga-substituted samples (small radii).  The JT-like
bands are not observed in the Cu-substituted compounds, indicating a weaker
oxygen disorder than in the Zn and Sc series, although the substitutional ions
have comparable radii.  

It is possible that the distortion of the oxygen sublattice results from two
distinct steps: first, the oxygens neighboring to the substitutional ions are
displaced directly by the stress produced by the substituent. Secondly, this
displacement can lead to a localization of the $e_g$ electrons around the
substitutional ion, stimulating an additional long-range distortion of the
oxygen sublattice by the JT effect.  This additional distortion can be enhanced
by electronic and magnetic properties of the substituent, e.g., the
impossibility of acquiring the $e_g$ electron in the case of 3$d^{10}$ electron
configurations of the substituent, a magnetic state of the substituent that
weakens the ferromagnetism, etc.  The suggestion of a long-range distortion
involving oxygens farther from the substituent is supported by the fact that
the frequency and the width of the broad bands does not differ between Sc- and
Zn-substituted compounds and \lamno\ [see Fig.~\ref{Zn}(c)]. This is in contrast
with the local vibrational modes (see Sec.~\ref{Aglike}) that depend strongly
on the type of substituent.  The secondary distortion can be the reason why the
Cu-substituted series differs from the Zn- and Sc-substituted ones. It is
possible that the electronic and magnetic structure of Cu ions does not support
the localization of the $e_g$ electrons and the related second step of the
distortions.  We discuss further this topic in Sec.~\ref{infra}.

The JT-like bands of Zn- and Sc-substituted samples persist in the
ferromagnetic phase (see Fig.~\ref{prehled10}) since they originate primarily
from the distortions of the oxygen sublattice induced by the substituents.
This trend is in contrast with the observations, e.g., in
La$_{0.8}$Sr$_{0.2}$MnO$_3$~\cite{Bjornson}, where the broad bands are caused
by JT distortions and both the JT distortions and the broad bands are
suppressed in the ferromagnetic phase.  Actually, the JT-like bands are more
pronounced at 77\K\ than at 300\K\ [see Fig.~\ref{Zn}(b)]. This is probably due
to a reduction of the width of the bands resulting from an increase of the
lifetime of the corresponding vibrational states with decreasing
temperature~\cite{AbrashevPhysStatSol}.

The broad bands are also present in the \lamno\, sample [see Fig.~\ref{Zn}(c)].
This sample is doped by vacancies and contains 35\% of Mn$^{4+}$ ions. The high
concentration of vacancies at the Mn site (5.1\%),  however,  transforms the
low temperature phase into a resistive noncollinear ferromagnet~\cite{Vergara}.
Consequently, the distortions of the oxygen sublattice and the broad bands persist
also below \tc. The broad bands with similar temperature evolution were
observed also in a La$_{0.7}$Ca$_{0.3}$MnO$_{3}$ sample~\cite{Pantoja}.

\subsubsection{The additional $A_{g}$-like high-frequency mode} \label{Aglike}
Let us discuss the peak observed  at the highest  frequencies of 640--$760\cm$.
Its common characteristics are (a) it appears only in Mn-substituted samples
and its intensity increases with the concentration of substituents (the
increase is most pronounced in the Cr and Co series), (b) it has $A_{g}$-like
properties: it is absent or strongly suppressed in the \CP\ spectra, (c) in
contrast to the JT-like bands, it has a relatively small spectral width, (d)
its frequency depends rather strongly on the type of the substituent.  We
assign the peak to a local breathing (in-phase stretching) mode of oxygens in a
close vicinity of the substituent ion.  This interpretation is further
supported by the symmetry of the mode: the suggested eigenvector corresponds to
a nondegenerate vibration, i.e., it has the $A_g$-like symmetry.  Moreover,
these modes are not infrared active (see Sec.~\ref{infra}) which implies that
the eigenvector should have a center of symmetry. The property (c) indicates
that the peak does not originate from the phonon density of states.  Filho
\etal~\cite{Filho} observed a similar peak at about $707\cm$ in
La$_{0.70}$Sr$_{0.30}$Mn$_{1-x}$Fe$_x$O$_3$ and interpreted it as the silent
$A_{2g}$ mode activated by a symmetry breaking or as the phonon density of
state feature.  However, this interpretation is in contradiction with the
property (d).

\begin{figure}[!t] 
\vspace{-5mm}
	\includegraphics[width=7cm]{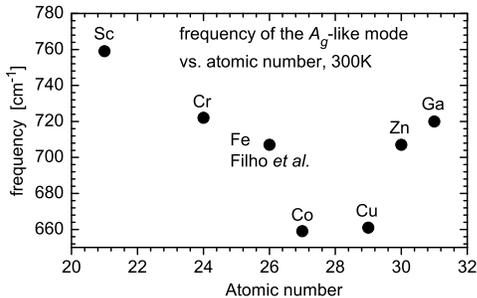}
\vspace{-5mm}
	\caption{The frequency of the additional $A_g$-like mode as a function of 
the atomic number.}  
\label{aglikefreq} 
\end{figure}

Figure~\ref{aglikefreq} summarizes the frequencies of the additional
$A_{g}$-like mode including the Fe data of Ref.~\cite{Filho}.  Evidently, the
frequency does not exhibit a simple mass effect. This is not surprising,
because the substitutional ions do not take part in the vibration.  The data do
not indicate a simple monotonic relation between the frequency and the \bo\
distance.  Some other factor than the radius of the substituents must play an
important role.  It could be, e.g., the degree of covalency of the {\it M}--O
bond, which  would imply a dependence of the frequency on the position of {\it
M} atom in the Periodic Table. Indeed, it can be seen in Fig.~\ref{aglikefreq},
that the frequency decreases when going from the borders of the group of
transition metals, where the covalency can be supposed to be small, towards the
center, where it should be larger.  Surprisingly, the intensity of the peak
exhibits the same trend: in Co-substituted compounds the intensity is the
largest and it is very small in  Sc- and Ga-substituted compounds. The peak of
the Fe-substituted compound observed in the data of Filho \etal~\cite{Filho} has
also a very large intensity as compared with the other phonons.  Another factor
affecting this frequency can be the valence of the substituent that influence
electrostatically the stiffness of the {\it M}--O$^{2-}$ bond.  This effect
would lower the frequencies corresponding to the substituents with valencies 2+
(Cu and Zn) with respect to their hypothetical 3+ state.  However, a clear
understanding of the frequency and intensity variations remains a challenge for
future studies. 

The properties of the high-frequency mode are even more complex. In the Co
series, the mode anomalously softens by about $15\cm$ with decreasing
temperature, and becomes rather asymmetric at 77\,K. It is possible that this
anomaly is due to the transition from the intermediate spin state occurring at
room temperature, to the low spin state at low temperature~\cite{Radaelli}.
These two states differ in electronic configuration as well as in ionic radius,
which could influence the frequency as discussed above.  The frequency
increases steeply in the temperature range between 200 and 300\K\ (see the inset
of Fig.~\ref{Co}) and we expect that it will continue to increase considerably
above 300\K, where Co passes into the high spin state~\cite{Radaelli}.  In the
Cu-substituted samples with $x=0.04$, 0.06, 0.08, the mode appears to consist
of a peak at $660$--$670\cm$ present both at room and at low temperature, and a
peak at about $690$--$700\cm$ emerging at low temperature. The two frequencies
are likely to correspond to two types of Cu--O bonds with different stiffnesses.
The difference in bonding can be caused, e.g., by the reduction of the cubic
symmetry of the CuO$_6$ octahedra due to the JT effect.  The frequency of the
mode of Zn- and Cu-substituted samples with $x=0.10$ is considerably lower (about
$9\cm$) than that of the $x=0.08$ samples.  This can be a consequence of an
interaction between the areas around the substituent ions, that must increase
with increasing concentration.

\subsubsection{The $A_{1g}$ soft mode} \label{soft} The $A_{1g}$ mode of the
rhombohedral structure has the frequency within the interval $190$--$220\cm$.
The vibrational pattern of this mode has the shape of the rhombohedral
distortion~\cite{Abrashev} (static rotational displacement of the oxygen
octahedra around the cubic [111] direction).  With increasing temperature
towards the transition from rhombohedral to cubic structure at about
$800\,^\circ$C~\cite{Carron2}, its frequency should decrease to zero, i.e., it is
a soft mode.  In our spectra, in accord with this interpretation, the $A_{1g}$
mode hardens when going from 300 to 77\K. In the parent compound \lsmo, the
hardening is about $12\cm$ in good agreement with the results of
Ref.~\cite{Podobedov}; in the spectra of the substituted samples, the frequency
change is weaker.  As noted by Abrashev \etal~\cite{Abrashev}, based on the
form of the eigenvector of the $A_{1g}$ mode, its frequency should mainly
correlate with the angle $\alpha$ of the rhombohedral distortion.  The angle
can be calculated from the coordinate $x_{\rm O}$ of the oxygens [($x_{\rm O}$,
0, 1/4) in a hexagonal setting] by using this equation~\cite{Chen} 
\begin{equation} x_{\rm
	O}=\frac{1}{2}\left(1\pm\frac{1}{\sqrt{3}}\tan{\alpha}\right)\;.
\end{equation}

\begin{figure}[!t]
\vspace{-5mm}
	\includegraphics[width=8.6cm]{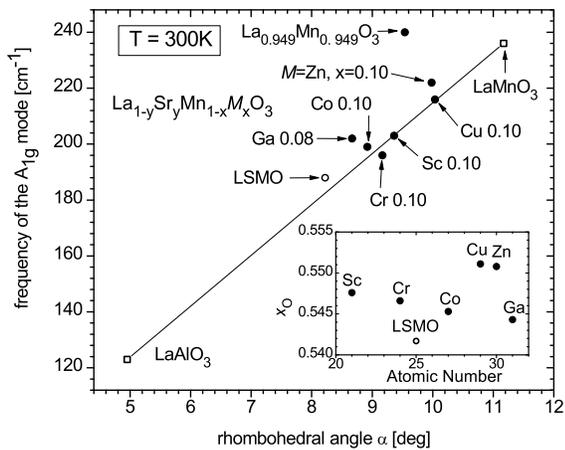}
\vspace{-10mm}
	\caption{Frequency of the $A_{1g}$ soft mode versus the angle of
rhombohedral distortion $\alpha$ of the \lsmbo\ samples and the parent $x=0$ (LSMO) sample. The data for LaMnO$_3$ and LaAlO$_3$ taken from
Ref.~\cite{Abrashev} are connected by the straight line.
The inset shows the values of the coordinate $x_{\rm O}$ of the oxygens versus 
the atomic number (Ref.~\cite{Fadli}).  
}
\label{alpha} \end{figure}

Figure~\ref{alpha} shows the dependence of the frequency of the $A_{1g}$ mode
on $\alpha$, the values of the latter quantity being calculated from those of
the $x_{\rm O}$ (Ref.~\cite{Fadli}) (see the inset of Fig.~\ref{alpha}).  The
frequency is almost a linear function of $\alpha$, and falls on the line
connecting the data for LaMnO$_3$ and LaAlO$_3$ of Ref.~\cite{Abrashev}.  The
angle $\alpha$ is influenced mainly by the Sr content $y$, putting the Cu and
Zn points apart from the other compounds.  The radius of the substituted ions
seems to play only a minor role.  For example, the values of the frequency and
$\alpha$ for Sc and Co, Cr, Ga are very similar, in spite of a significantly
larger radius of Sc compared to the remaining ions.  The only outlier is the
\lamno\ ($x_{\rm O}=0.5485\,{\rm \AA}$ taken from Ref.~\cite{Boix}).  Since it
is the only compound among our samples that has a significant concentration of
vacancies, we attribute the deviation to this structural difference.  The
linear dependence of the frequency of the octahedra tilt mode on the tilt angle
has been found also in orthorhombic manganites~\cite{Carron}.

\subsection{Infrared reflectance} 
\label{infra} 
\begin{figure}[!t]
\vspace{-5mm}
	\includegraphics[width=8.6cm]{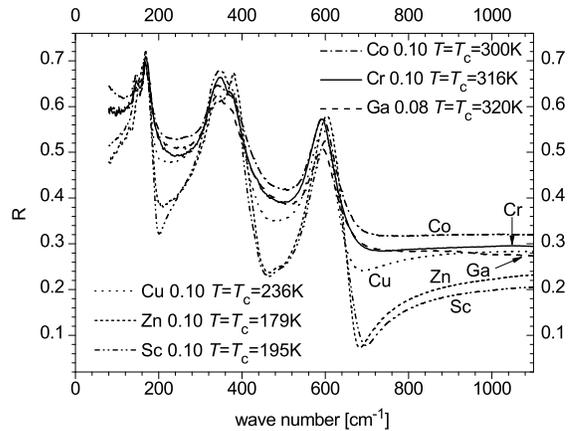}
\vspace{-10mm}
	\caption{Infrared reflectivities of the \lsmbo,  $x=0.10$ compounds (except for $M=$ Ga with $x=0.08$)
measured  at the critical temperatures of each sample.}
\label{ReflData}
\end{figure}

\begin{figure*}[!t]
\vspace*{-10mm}
	\includegraphics[width=16cm]{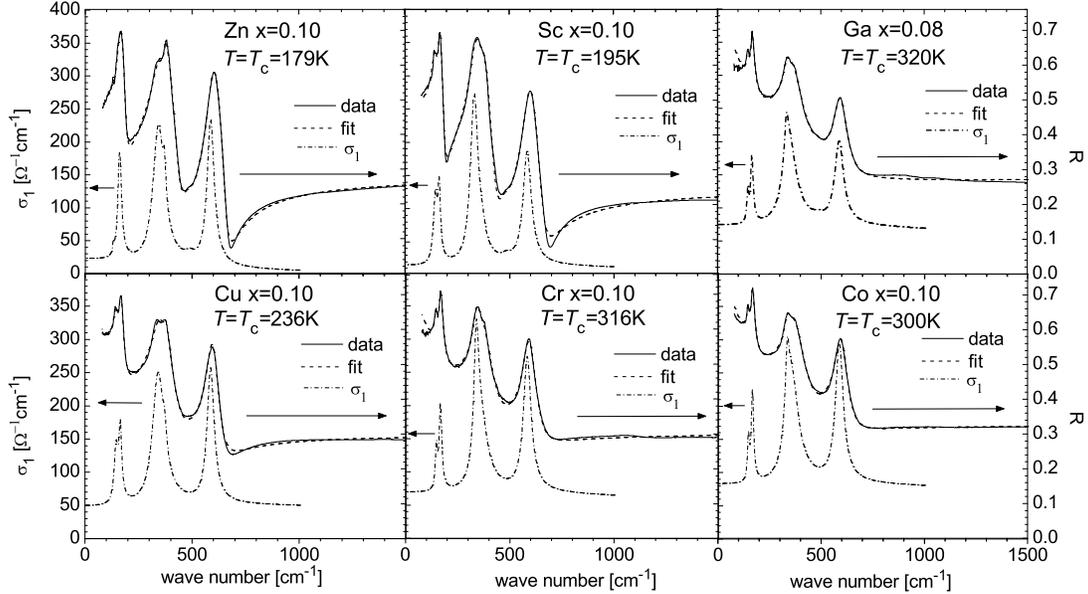}
\vspace{-10mm}
\caption{Infrared reflectivities of Fig.~\ref{ReflData} (solid lines) with fits
	(dashed lines) and the real part of complex conductivity (dash-dotted lines). The
	reflectivities and the conductivities correspond to the right and left axes,
	respectively.}
\label{ReflFity}
\end{figure*}

Figure \ref{ReflData} presents the measured infrared reflectivities of the
$x=0.10$ samples (including the Ga-substituted sample with $x=0.08$).  In order
to compare the samples in the same magnetic state, the spectra were measured at
the critical temperature of each compound.   In the region from 130 to $600\cm$
the spectra exhibit several phonon structures reported already in earlier
studies~\cite{MarziInfra}.  No phonon modes are observed (within the noise
$\leq0.1\%$) in the frequency region of the additional $A_g$-like Raman lines
(640--$800\cm$). This excludes the possibility that the Raman lines are due to
infrared phonons activated by a reduction of symmetry. In the following, we
focus on the long-wavelength behavior of the itinerant electrons, and extract
the dc  resistivity.  We recall that the dc resistivity of a polycrystal
determined from the infrared spectroscopy is likely to be closer to the
intrinsic resistivity than the data obtained by standard transport measurements
because it is much less sensitive to the intergrain boundaries~\cite{Kim}.
Already the raw reflectivity data of the Zn- and Sc-substituted samples can be
seen to differ from the others by a lower background and more pronounced phonon
structures, which corresponds to a higher dc  resistivity.  In order to explore
the difference quantitatively, we have fitted the spectra using the model
dielectric function in the standard form of a  sum of Lorentzian oscillators, 
\begin{equation} 
\label{DielectricFunction}
\epsilon(\omega)=\epsilon_\infty-\frac{F_{\rm D}}{\omega(\omega+{\rm
i}\gamma_{\rm D})} +\sum_j\frac{F_j}{\omega_j^2-\omega^2-{\rm i}\omega\gamma_j}\;, 
\end{equation}
where $\epsilon_\infty$ stands for the electronic interband contribution, the
second term is the contribution of the conduction electrons, and the sum
represents phonons.  Normal incidence reflectivity was calculated as
$R=|(1-N)/(1+N)|^2; N=\sqrt{\epsilon}.$ We have fitted the data in the interval
80--$1500\cm$.  This region is sufficient to yield a reasonable description of
the behavior of the conducting electrons, keeping a simple model with a small
number of parameters, which would be impossible when including higher
frequencies.  The fitted reflectance spectra and the resulting real part of the
complex conductivity, $\sigma(\omega)=-{\rm
i}\omega\epsilon_0\epsilon(\omega)$, are plotted in Fig.~\ref{ReflFity}.  The
agreement between the data and fits is reasonable; resulting values of the
parameters are shown in Table~\ref{TablePar}.

\begin{table}[!t] 
\begin{tabular}{rrrrrrr}
\hline
\hline
\vrule height 10pt width 0cm
& Zn 10\%& Sc 10\%&  Ga 8\%& Cu 10\%& Cr 10\%& Co 10\%\\ 
\hline
\vrule height 10pt width 0cm
$\epsilon_\infty$& 10.2& 8.7& 11.3& 12.1& 12.9& 14.5\\ 
$F_{\rm D}$& 	576 & 1400& 13400& 16200& 12600& 19600\\ 
$\gamma_{\rm D}$& 411& 1505& 3000& 5380& 3000& 3917\\ 
$F_1$& 		5& 146& 29& 112& 41& 42\\ 
$\omega_1$& 	131& 146& 146& 145& 147& 147\\ 
$\gamma_1$& 	6& 23& 10& 22& 11& 11\\ 
$F_2$& 		204& 82& 93& 100& 118& 113\\ 
$\omega_2$& 	160& 164& 166& 164& 167& 166\\ 
$\gamma_2$& 	20& 13& 15& 16& 15& 13\\ 
$F_3$& 		784& 686& 66& 765& 323& 140\\ 
$\omega_3$& 	341& 332& 333& 340& 339& 335\\ 
$\gamma_3$& 	63& 45& 20& 65& 28& 23\\ 
$F_4$& 		53& 51& 615& 46& 438& 726\\ 
$\omega_4$& 	370& 370& 346& 371& 362& 357\\ 
$\gamma_4$& 	15& 40& 80& 21& 75& 77\\ 
$F_5$& 		25& 20& 15& & & \\
$\omega_5$& 	488& 487& 485& & & \\
$\gamma_5$& 	50& 50& 50& & & \\
$F_6$& 		540& 488& 381& 488& 543& 571\\ 
$\omega_6$& 	586& 586& 587& 585& 583& 587\\ 
$\gamma_6$& 	40& 48& 51& 40& 45& 46
\vrule height 10pt width 0cm \\
\hline
\hline
\end{tabular}
	\caption{Best-fit values of the parameters of the dielectric
	function~(\ref{DielectricFunction}). The units of
	$F_j$ are $10^3\,\rm cm^{-2}$, the units of $\omega_j$ and $\gamma_j$ are $\cm$.  
	The error of the dc resistivity $\rho(\omega=0)=
	\gamma_{\rm D}/ (\epsilon_0F_{\rm D})$ is about 10\%, however, the errors of $F_{\rm D}$ and
	$\gamma_{\rm D}$ itself are large due to the flat conductivity background in the
	FIR region.  The errors of the remaining values are a few percent except for the
	widths and strengths of the overlapping phonons at $340$ and $370\cm$, where the
	uncertainty is about 50\%.} 
\label{TablePar} 
\end{table}

\begin{figure}[!t] 
\vspace{-5mm}
	\includegraphics[width=7cm]{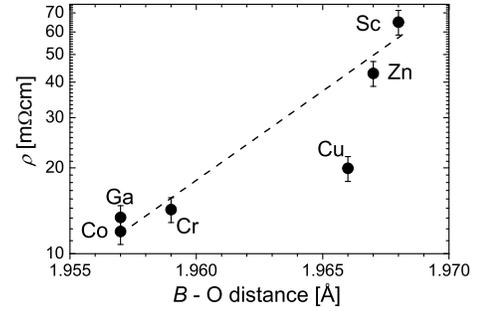}
\vspace{-5mm}
	\caption{The values of the dc resistivity of the \lsmbo,  $x=0.10$ compounds (except for $M=$ Ga with $x=0.08$)
extrapolated from the infrared data. The dashed line is shown to guide the eye.}  
\label{resistivity} 
\end{figure}

Figure~\ref{resistivity} shows the values of the dc resistivity, $\rho=1/{\rm
Re\,}\sigma(\omega=0)$.  The trend of increasing resistivity with increasing
\bo\ distance is clearly visible; the values corresponding to the large
substituents, Sc and Zn, are significantly larger than those of the smaller
ones.  The resistivity of the Cu-substituted compound is close to that of the
group of smaller substituents. This is the same trend as indicated by the
JT-like bands in the Raman data. 

We explain the difference between the Cu- and Zn-substituted compound in terms
of the occupancy of the $3d$ orbitals as follows: the Zn$^{2+}$ ion has fully
occupied $3d$ orbitals in contrast with the Cu$^{2+}$ ion, that has one $3d$
orbital empty. This orbital can acquire a conduction electron, and creates a
nonzero magnetic moment that might enhance the ferromagnetism.  However, the
Ga$^{3+}$ ion has the $3d$ shell occupied and has no magnetic moment, but the
samples belong to the low resistivity group. The Ga substitution shows that the
structural and electric properties of the compounds are mainly related to the
substituent radius, and can be partially influenced by its electronic and
magnetic structure; this seems to be the case in the Cu-substituted compound.

Alonso~\etal~\cite{Alonso} suggested that an important factor influencing
magnetic properties of the substituted manganites is the electrostatic
potential created by the substituent.  In this picture, because the average
valence of the Mn ions is 3.3+, holes are attracted to the substituent with a
lower valence (3+ or 2+), and, consequently, the holes tend to localize and
\tc\ decreases.  If this effect were important, the \tc\ of the
Zn$^{2+}$-substituted sample would be significantly lower than that of the
Sc$^{3+}$-substituted compound, because the valence 2+ should localize holes
more strongly than the valence 3+.  In our samples, however, the \tc's of these
compounds are nearly the same.  This fact indicates that the influence of the
electrostatic potential on the localization of holes is significantly less
important than that of the local pressure.

Ghosh \etal~\cite{Ghosh} studied the effects of Mn substitutions by
transition elements in La$_{0.7}$Ca$_{0.3}$MnO$_3$.  It has been found that the
changes of the maximum of magnetoresistance correlate with those of the lattice
parameter, and that \tc\ correlates with the conductivity at \tc.  The latter
result is basically confirmed by our studies.

\section{Summary and Conclusions} \label{conclusion} We have used Raman and
infrared spectroscopy, magnetic measurements, and x-ray analysis to study the
influence of Mn-site substitutions in optimally doped manganites.  Our results
support the idea that the main factor influencing the oxygen disorder and
electric and magnetic properties of these compounds is the radius of the
substituents and that their electronic and magnetic structure plays only a
minor role.  Our substituents form two groups, with the radii significantly
larger than (Zn$^{2+}$ and Sc$^{3+}$), and similar to (Co$^{3+}$, Cr$^{3+}$,
Ga$^{3+}$), the average radius of the Mn ions.  The former group differs from
the latter by the occurrence of JT-like bands in the Raman spectra, reflecting
an enhanced oxygen disorder, higher values of dc resistivity, and lower values
of \tc.  This is in agreement with the picture proposed in the work of
Ghosh~\etal~\cite{Ghosh} on La$_{1-y}$Ca$_y$Mn$_{1-x}$$M_x$O$_3$, suggesting
that the local stress produced by the substitutional ion modifies the Mn--O
distances around the substitutional ion.  This leads to changes of the energies
of the $e_g$ orbitals, and to a localization of the itinerant electrons which
competes with the DE interaction.  Consequently, the JT disorder is enhanced,
dc resistivity increases, and \tc\ decreases. These trends are weaker in the
case of Cu substitution although the ionic radius is close to that of Zn and
Sc.  It is possible that some additional effects due to the electronic or
magnetic structure of Cu enhance the mobility of electrons. 

The Raman spectra exhibit an additional $A_g$-like mode that we attribute to
the local breathing mode of oxygens in a close vicinity of the substituent ion.
Its frequency and intensity are very sensitive to the type of substituent. They
seem to depend on the position of the substituent ion in the Periodic Table
suggesting that the electronic characteristics of the \mo\ bond are important.
However, the origin of these effects remains an open question for future
studies.  The mode of the Co-substituted samples exhibits an anomalous
softening of $15\cm$ when going from 300 to 77\K, which is likely due to the
spin state transition of Co. 

We believe that our results may be helpful
for understanding the electronic properties of
manganites, in particular, 
and transition metal oxides, in general.\\

\begin{acknowledgments} A.D. acknowledges discussions and help of N.~Garro,
	V.~K\v{r}\'apek, D.~Munzar and S.~Valenda.  The work has been supported by 
	the UE through a TMR Grant (No. HPRNCT2000-00021) and 
	Project No. MSM 00216 22410 of the Ministry of Education of Czech Republic.
\end{acknowledgments}

\newpage 

\end{document}